\documentclass[preprint,preprintnumbers,amsmath,amssymb]{revtex4}
\usepackage{graphicx}
\usepackage{dcolumn}
\usepackage{bm}

\begin{document}

\title{Quantum thermodynamics of ergotopy for a relativistic battery as a witness to Unruh-Hawking thermality in curved (A)dS spacetimes}
\author{Xiang Hao}
\altaffiliation{Corresponding author}
\email{xhao@mail.usts.edu.cn}
\author{Tao-Feng Gan}
\author{Cheng-Tai Wu}
\author{Tian-Xi Ren}
\author{Wei-Wei Zhang}
\author{Yin-Zhong Wu}
\affiliation{School of Physical Science and Technology, Suzhou University of Science and Technology, Suzhou, Jiangsu 215009, People's Republic of China}
\affiliation{Pacific Institute of Theoretical Physics, Department of Physics and Astronomy,
\\University of British Columbia, 6224 Agriculture Rd., Vancouver B.C., Canada V6T 1Z1.}

\begin{abstract}

We propose a relativistic quantum battery model consisting of an accelerated Unruh-DeWitt detector coupled to a massless scalar field in a de Sitter and anti-de Sitter  spacetimes. The maximal amount of quantum extractable work, defined as the ergotropy, is used to probe Unruh-Hawking thermality induced by acceleration and spacetime curvature. Using the open quantum system approach, we study the dynamics of the ergotropy with respect to the Kubo-Martin-Schwinger temperature, spacetime boundary conditions, and dimensionality. It has been found that the asymptotic value of quantum work extraction is determined solely by the acceleration and curvature. The steady behavior in the two spacetimes can be unified to witness the global thermality which is independent of boundary conditions and dimensionality. From a local perspective, we investigate how the ergotropy evolves through different pathways as the battery gradually reaches the same thermal equilibrium state characterized by a certain KMS temperature. In dS spacetime, the evolution at large acceleration exhibits pronounced oscillations and differs from the fast thermal relaxation observed at low acceleration. Varying the boundary condition in AdS spacetime can improve the energy storage of the moving battery. When the dimension of AdS spacetime is increased, vacuum fluctuations can modestly amplify the ergotropy in the initial stage and facilitate the rapid thermalization. From the perspective of energy transfer, the relativistic quantum battery helps explore the thermal vacuum in curved spacetimes.

\end{abstract}

\maketitle

\section{Introduction}

The unification of quantum theory and general relativity continues to be one of the most significant unresolved challenges in physics today. In recent years, a potential valuable approach to understanding quantum gravity has been that of relativistic quantum information \cite{Mann2012}, which aims to explore both the effects of relativity on quantum information protocols and the ways quantum information resources can be used to probe field quantization in various spacetime backgrounds. As solutions to Einstein's field equations in general relativity, de Sitter (dS) spacetime, anti-de Sitter (AdS) spacetime, and flat Minkowski spacetime are known as the three maximally symmetric spacetimes of constant curvature. The difference in vacuum structure between curved and flat spacetimes gives rise to distinct radiation reactions in the Kubo-Martin-Schwinger (KMS) condition \cite{Unruh1976,Hawking1975,GibbonsHawking1977,Fulling1973,Davies1975}. The notions of a vacuum state and particle number become observer-dependent in curved spacetimes. The evidence was demonstrated by the Unruh effect \cite{Unruh1976} and Hawking effect \cite{Hawking1975,GibbonsHawking1977}. For accelerated observers in a flat Minkowski spacetime, the vacuum state is regarded as a thermal state, with a certain Unruh temperature proportional to their proper acceleration. Although the Unruh effect was initially discovered in a flat Minkowski spacetime, this phenomenon has been extended to different curved spacetimes. In dS, an accelerated detector experiences a thermal response at a Gibbons-Hawking temperature dependent on the acceleration and curvature \cite{Deser1997,Casadio2011}. In contrast, AdS with a negative curvature takes on the Unruh thermality only if the acceleration of the detector exceeds the curvature scale of the spacetime \cite{Avis1978,Jennings2010}. The task of discriminating between the vacuum structures of different spacetimes becomes more challenging as the dimensionality of spacetime increases \cite{Takagi1986,Ooguri1986,Louko2016}.

The general thermalization theory shows that the KMS condition holds for the global Rindler horizon and presents the loss of information in a flat spacetime \cite{Crispino2008}. However, the KMS condition does not determine the particular shape of the response spectrum for an accelerated observer in curved spacetimes. In addition, the inversion of statistics occurs in odd-dimension Minkowski spacetime, where a Bosonic (Fermionic) field exhibits a Fermi-Dirac (Bose-Einstein) distribution \cite{Louko2016}. The spurious statistic inversion was ascribed to the absence of Huygens principle \cite{Takagi1986,Ooguri1986} or a local feature of the accelerating detector \cite{Sriramkumar2003,Arrechea2021}. Recently, a relativistic quantum information approach has provided a novel means of estimating the global characteristics of Unruh thermality. Some quantum resources \cite{Yu2012,Doukas2013,Martin2015,Feng2018,Huang2018}and methods from quantum metrology \cite{Hao2019,Du2021,Chowdhury2022,Feng2022} have been employed to detect the Unruh effect using the Unruh-DeWitt (UDW) detector model. This pointlike two-level detector moves along the world line depicted as a trajectory $x(\tau)$, where $\tau$ is the proper time of the UDW detector. The interaction between the detector and scalar field in spacetimes will result in the excitations of the detector. The thermal response of the UDW detector can be used to characterize vacuum structure in a flat spacetime. Recent studies have employed uncertainty relations \cite{Li2025}, quantum coherence \cite{Zhang2022}, geometric phase \cite{Hu2012}, and Fisher information \cite{Feng2018,Du2021} as effective probes for Unruh thermality. Thus, a question naturally arises that how to develop an operational way to probing the thermalization in curved spacetime like dS and AdS. Witnessing Unruh thermality of curved spacetimes plays a key role in understanding spacetime structures. The extension of relativistic quantum information approaches to curved spacetimes is a natural progression.

We put forward a relativistic UDW battery and exploit quantum work extraction to examine Unruh-Hawking thermality of curved spacetimes including dS and AdS. From a viewpoint of quantum information apparatus, quantum battery is viewed as quantum system where quantum resources can be used to store energy from an external field or quantum chargers \cite{Alicki2013,Campaioli2017,Le2018,Ferraro2018,Zhang2019,GarciaPintos2020,Farina2019,Ghosh2021,Cruz2022,Barra2022,Rossini2020,Hao2023,Zhang2025,Chen2025}. For a quantum battery in an active state, quantum work extraction can be performed by cyclically unitary operations. The ergotropy, defined as the maximum amount of work that can be extracted from a quantum system, serves as a fundamental quantity for assessing the charging performance of a quantum battery. In particular, we study the dynamics of the ergotropy for the UDW battery coupled to a massless scalar field in dS and AdS spacetime. By means of the open quantum system approach, we can obtain the thermodynamic features of quantum work extraction determined by the response function. In the asymptotic condition, the steady ergotropy can demonstrate the global side of Unruh thermality related to the curvature. During the transient stage, the specific dynamical behaviors exhibit diverse characteristics across different curved spacetimes, with changes in spacetime dimensions and boundary conditions. It is interesting to explore the interplay between quantum work extraction and vacuum structure from the viewpoint of relativistic quantum thermodynamics. The relativistic UDW battery will offer deeper insights into the thermal nature and vacuum structure of curved spacetimes.

What motivates the exploration of thermal vacuum for curved spacetimes via quantum work extraction is the following factors. First, quantum work extraction operationally estimates energy transfer between the battery and the environment, thereby encoding traits of the response function, which is closely tied to the vacuum structure of curved spacetimes. Thermal states arise when event horizons are present. Quantum work extraction serves not merely as an energy storage task but as a functional probe of vacuum fluctuations and Unruh thermality. The relativistic battery provides a new way to observe the global side of the Unruh effect in different curved spacetimes. In addition, distinct evolutions of quantum work extraction can reveal the evidence of quantum nature of vacuum fluctuations, specifically manifesting as quantumness across the Rindler horizon \cite{Ahmadi2014,Zhao2020,Liu2021}. Crucially, the decay of quantumness in relativistic quantum batteries originates from field fluctuations in curved spacetimes. To this end, we construct an accelerated quantum battery coupled to a massless vacuum scalar field in the curved  spacetimes like dS and AdS, treating the relativistic battery as an open quantum system and the fluctuating vacuum as its environment. Therefore, by analyzing how quantum work extraction responds to changes in spacetime dimensions and boundary conditions, we can provide an operational interpretation for Unruh-Hawking thermality.

In this paper, we carry out an investigation of relativistic quantum battery in curved spacetimes, restricting our attention to the constant-curvature dS and AdS cases. We make use of quantum work extraction as a probe of Unruh-Hawking thermality in curved spacetimes. We will be working in Plank units with $c=\hbar=G=1$ throughout the paper. The paper is organized as follows. In Sec. II, we introduce the general formalism of the ergotropy for the UDW battery. Subsequently, we give an overview of scalar fields in dS and AdS spacetimes. In Sec. III, by using the open system approach, we explore how the response function governs the thermodynamics of quantum work extraction in dS and AdS spacetimes. In Sec. IV, we study the evolutions of the ergotropy, which will consist of the asymptotic behavior and the transient dynamics. We compare the probing results for the two curved spacetimes. For AdS spacetime, the effects of boundary conditions and dimensionality on the ergotropy are studied in detail. Finally, in Sec. V, we give our conclusions and discussions.

\section{General formalism}

\subsection{Erogtropy for the UDW battery}

We put forwad a proposal of an accelerated UDW battery in curved spacetime. The model of UDW battery consists of a two-level atom moving along a uniformly accelerated trajectory. The static battery is generated by the Hamiltonian of $H_0=\omega_0 \sigma^{+}\sigma^{-}$ where $\omega_0$ denotes the transition frequency between an excited state $|e\rangle$ and a ground state $|g\rangle$. $\sigma^{\pm}$ represent the rising and lowing operator respectively. To realize a charging process, we need apply a classical coherent field to drive the battery. The external field with the frequency $\omega$ is described as $\mathbf{E}(t) = {\mathbf{E}_0}\cos \omega t $, where ${\mathbf{E}_0} $ represents the amplitude of the field. The Hamiltonian of the driven battery can be written as $\tilde{H} = H_0 - \mathbf{d}\cdot \mathbf{E}$ where the second term denotes the dipole interaction between the external field and battery. The parameter $\mathbf{d}=\mathbf{d}_0(\sigma^+ +\sigma^-)$ represents the electric dipole moment of a two-level atom. In the interaction picture, the transformed Hamiltonian $H_b=e^{iH_0 t} \tilde{H} e^{-iH_0 t}$ can be given by $H_b=\frac {\Omega}2[\sigma ^+(e^{i(\omega_0+\omega) t} +e^{i(\omega_0-\omega) t} )+ \sigma ^- (e^{-i(\omega_0+\omega) t}+e^{-i(\omega_0-\omega) t})]$ where we can neglect the high frequency oscillating parts $e^{\pm i(\omega_0+\omega)t}$ in the rotating wave approximation. Here $\Omega=-\mathbf{d}_0\cdot \mathbf{E}_0$ represents the effective coupling strength. Considering the resonance condition of $\omega=\omega_0$, we can obtain the expression of ${H_b} = \mu(t)\frac {\Omega}{2}({\sigma ^+} + {\sigma ^-} )$ where $\mu(t)=1(0\leq t \leq \tau)$ is the switching function for the charging process.

At $\tau=0$, the battery is prepared in the ground state, which describes the state of the depleted battery. In the condition of no acceleration, the evolved state of the inertial battery can be governed by $\rho(\tau)=U(\tau)\rho_0 U^{\dag}(\tau)$ where the cyclical unitary operator is $U(\tau)=\mathcal{T}\exp [-i\int_{0}^{\tau} \mathrm{d}s H^{(b)}(s) ]$. Here, the symbol $\mathcal{T}$ denotes the time ordering operator and $\rho_0$ is an initial state. According to \cite{Alicki2013}, the maximal amount of quantum work extraction over all unitary transformations $\{ U(\tau)  \}$, i.e., the ergotropy, can be written as
\begin{equation}
\label{eq:(1)}
\mathcal{W}(\tau)=\mathrm{Tr}[H_0\rho(\tau)]-\min_{\{ U \}}\mathrm{Tr}[U\rho(\tau)U^{\dag}H_0].
\end{equation}
The minimal unitary transformation $U_{\sigma}$ is optimized to transform the battery into a passive state where no work can be extracted. The passive state satisfies that  $U_{\sigma}\rho(\tau)U^{\dag}_{\sigma}=\sum_{j}\varrho_j|\epsilon_j\rangle  \langle \epsilon_j |$ where $|\epsilon_j\rangle $ is the energy level state of $H_0$ with the corresponding energy $\epsilon_j$ in the increasing order and the eigenvalues $\varrho_j$ of $\rho(\tau)$ are arranged in the decreasing order. Therefore, the simplified expression of the ergotropy is given in the form of
\begin{equation}
\label{eq:(2)}
\mathcal{W}(\tau)=\sum_{j,k}\varrho_j \epsilon_k (|\langle \varrho_j|\epsilon_k\rangle|^2-\delta_{jk}).
\end{equation}
The larger maximal amount of extractable work, the better charging performance of quantum battery.

For a two-level atom, the density matrix  $\rho(\tau)$ can be expressed in the form of
$\rho(\tau)=\frac {I+\sum_j r_j(\tau)\sigma_j}{2}$ where $r_j=\mathrm{Tr}(\sigma_j \rho)$ is the $j$th-component of the Bloch vector. To obtain the ergotropy, we should firstly calculate the eigenvalues $\varrho_{1,2}=\textstyle{{1 \pm \left| {\mathbf{r}(\tau )} \right|} \over 2}$ of the density matrix of the battery. The optimal unitary operation is expressed as $U_\sigma =\sum\nolimits_{j=1,2} {\left|  \epsilon_j\right\rangle \left\langle \varrho_j \right|}$ where $\left|  \varrho_j\right\rangle$ is the eigenvector of the density matrix of the battery state. In the space of $\{|\epsilon_1\rangle=|g\rangle ,|\epsilon_2\rangle=|e\rangle\}$, $|\varrho_{1}\rangle=\sqrt{\frac {r+r_3}{2r}}|g\rangle+\frac {r_1+ir_2}{\sqrt{2r(r+r_3)}}|e\rangle$ and $|\varrho_{2}\rangle=\frac {r_1-ir_2}{\sqrt{2r(r+r_3)}}|g\rangle-\sqrt{\frac {r+r_3}{2r}}|e\rangle$ are obtained. The ergotropy can be achieved by the optimal unitary transformation,
\begin{equation}
\label{eq:(3)}
U_{\sigma}=\left( \begin{array}{cc} \sqrt{\dfrac {r+r_3}{2r}} & \dfrac {r_1-ir_2}{\sqrt{2r(r+r_3)}} \\ \dfrac {r_1+ir_2}{\sqrt{2r(r+r_3)}} & - \sqrt{\dfrac {r+r_3}{2r}}  \end{array}  \right),
\end{equation}
where $r=\left| {\mathbf{r}(\tau )} \right|$ represents the norm of the Bloch vector. The value of the ergotropy $\mathcal{W}$ is determined by the internal energy $E(\tau)=\mathrm{Tr}[H_0\rho(\tau)]=\frac {\omega_0(1+r_3)}2$ and the part of $\mathrm{Tr}[U_{\sigma}\rho(\tau)U_{\sigma}^{\dag}H_0]=\mathrm{Tr}[\rho_{\sigma}H_0]=\sum_{j=1,2}\varrho_j \epsilon_j= \frac {\omega_0(1-r)}2$. Therefore, the ergotropy is expressed as $\mathcal{W}=\frac {\omega_0(r+r_3)}2$. For convenience, we can define the scaled ergotropy as $\xi(\tau)=\frac {\mathcal{W}}{\omega_0}$,
\begin{equation}
\label{eq:(4)}
\xi(\tau)=\frac {1}{2}[r(\tau)+r_3(\tau)].
\end{equation}
In this protocol, a relativistic UDW battery is regarded as an open system which is coupled to a fluctuating vacuum field in curved spacetimes. The ergotropy of an open quantum battery will undergo quantum dissipation from the interaction between the battery and the vacuum scalar field. The total Hamiltonian of the UDW battery along an accelerated trajectory $x(\tau)$ in curved spacetime can be expressed as
\begin{equation}
\label{eq:(5)}
H=\mu(H_b+H_{\phi})+H_{I}.
\end{equation}
During the charging proper time $\tau$, the switching function $\mu$ is used to control the interplay between the battery and scalar field. The part $H_{\phi}=\sum_{\mathrm{k}}\omega_{\mathrm{k}}a^{\dag}_{\mathrm{k}}a_{\mathrm{k}}$ denotes the quantized Hamiltonian consisting of the massless scalar field $\Phi \left( {x(\tau )} \right)=\sum_{\mathrm{k}}[u_{\mathrm{k}}(x)a_{\mathrm{k}}+u_{\mathrm{k}}^{*}(x)a_{\mathrm{k}}^{\dag}]$ where the conditions of $\sum_{\mathrm{k}} u_{\mathrm{k}}(x)u_{\mathrm{k}'}^{*}(x)\propto \delta_{\mathrm{k}\mathrm{k}'}$ and $[a_{\mathrm{k}},a_{\mathrm{k}'}^{\dag}]=\delta_{\mathrm{k}\mathrm{k}'}$ are satisfied. The term $H_{I} =\lambda ({\sigma ^ + } + {\sigma ^ - })\Phi \left( {x(\tau )} \right)=\lambda( \sigma_1 \otimes \Phi )$ represents the interaction between the battery and scalar field. $\{ \sigma_{j},\;j=1,2,3 \}$ represent the three components of Pauli operator. In the condition of weak couplings, i. e., $\lambda \ll \Omega$, the UDW battery is charged by both the coherent driving field and the scalar field in curved spactimes. To further understand the interaction Hamiltonian, we will present the features of vacuum state in the two representative curved spacetimes.

\subsection{de Sitter and anti-de Sitter spacetime}

We will review scalar fields in both $\mathrm{dS}_4$ and $\mathrm{AdS}_4$ which are represented as $4-$dimensional hyperboloids embedded in $5-$dimensional spacetimes,
\begin{equation}
\label{eq:(6)}
ds^2=-dX_0^2+dX_1^2+dX_2^2+dX_3^2\pm dX_4^2.
\end{equation}
Here, the plus sign denotes $\mathrm{dS}_4$ and the negative one represents $\mathrm{AdS}_4$ spacetime. The well-known static (A)dS metric,
\begin{equation}
\label{eq:(7)}
ds^2=-\left( 1\pm \frac {r^2}{\ell^2} \right) dt^2+ \left( 1\pm \frac {r^2}{\ell^2} \right)^{-1} dr^2+r^2 d\Omega_2^2,
\end{equation}
is obtained via the coordinate transformations $X_1=r\sin\theta \sin \phi, X_2=\sin\theta \cos \phi, X_3=r \cos \theta$ and
\begin{align}
\label{eq:(8)}
X_0 &\;=\; \sqrt{\ell^2-r^2}\sinh (t/\ell),  X_4 \;=\; \sqrt{\ell^2-r^2}\cosh(t/\ell) \;\;\;\; \mathrm{for \;\;dS} \nonumber \\
X_0 &\;=\; \sqrt{\ell^2+r^2}\sin (t/\ell),\;\; X_4 \;=\; \sqrt{\ell^2+r^2}\cos(t/\ell) \;\;\;\; \mathrm{for \;\;AdS}
\end{align}
for $\mathrm{dS}_4$ and $\mathrm{AdS}_4$ respectively. The length $\ell=\sqrt{3/\Lambda}=1/k$ is referred to as the spacetime length. The dS case has a coordinate singularity at the cosmological horizon $r=\ell$. Here $k=\frac 1{\ell}$ denotes the curvature with the negative cosmological constant in AdS.

For $\mathrm{dS}_4$ we will also obtain the metric $ds^2=-dt^2+e^{2t/\ell}\sum_{j=1}^3 x_j^2$ by using comoving coordinates $(t,x_1,x_2,x_3)$ via the transformations
\begin{equation}
\label{eq:(9)}
\mathbb{T}\;=\;\ell \sinh(t/\ell)+\frac {r^2e^{t/\ell}}{2\ell}, X_0\;=\;\ell \cosh(t/\ell)-\frac {r^2e^{t/\ell}}{2\ell}, X_j\;=\;e^{t/\ell}x_j\;\;(j=1,2,3)
\end{equation}
We analyze a scalar field $\Phi$ via the action
\begin{equation}
\label{eq:(10)}
S=\int \sqrt{-g}\left[ \frac 12 g^{\mu\nu} \nabla_{\mu}\Phi \nabla_{\nu}\Phi-\frac 1{12}R\Phi^2 \right].
\end{equation}
The conformal vacuum is taken into account because it is the natural vacuum that obeys the symmetry of the de Sitter group \cite{Avis1978}. Therefore, the Wightman function $ G(x-x')=\langle 0|\Phi \big( x(\tau) \big)\Phi \big( x'(\tau^{'}) \big)|0\rangle$ for a comoving observer in the conformal vacuum of $\mathrm{dS}_4$ is given by
\begin{equation}
\label{eq:(11)}
G^{\mathrm{dS}}(x-x')=-\frac 1{2\pi \ell\sqrt{2}}\left[ \dfrac {1}{\sinh^2(\frac {\tau-\tau'}{\ell}-i\epsilon)}  \right].
\end{equation}
Here the trajectory is stationary for the reason that the above expression is a function of $\Delta \tau=\tau-\tau'$ only.

By quantizing the field in the Einstein Static Universe and then translating the result to AdS spacetime, we can obtain the Wightman function for the vacuum, which is written as,
\begin{equation}
\label{eq:(12)}
G^{\mathrm{AdS}}(x-x')=\frac 1{4\pi \ell\sqrt{2}}\left[ \dfrac {1}{\sigma(x,x')}-\frac{\zeta}{\sigma(x,x')+2}  \right],
\end{equation}
where $2\ell^2\sigma(x,x')$ is the geodesic distance between $x$ and $x'$, and the parameter $\zeta=-1,0,1$ represents three cases: Dirichlet ($\zeta=1$), transparent ($\zeta=0$), and Neumann ($\zeta=-1$) boundary conditions \cite{Deser1997,Jennings2010,Du2021}. The geodesic distance can be obtained by the embedding space.

We employ the UDW battery moving along a specific trajectory, where the thermal radiation can be perceived by an accelerated observer. The probability of energy transition from the ground state $|g\rangle$ to the excited one $|e\rangle$ is determined by the response per unit time,
\begin{equation}
\label{eq:(13)}
\mathcal{G}(\Omega)=\int_{-\infty}^{\infty} \mathrm{d}\Delta \tau \cdot e^{-i\Omega\Delta \tau} G(\Delta \tau),
\end{equation}
which is the Fourier transform of the Wightman function.

The key utility of the UDW battery is that the physical model can provide an operational way to interpreting the thermality, more generally, vacuum structures for different curved spacetimes. Thermal states occurs in the expanding de Sitter spacetime \cite{GibbonsHawking1977}, in the presence of a black hole \cite{Hawking1975}, and for uniform accelerations \cite{Unruh1976}. When the acceleration is sufficiently large, we demonstrate that the UDW battery will experience Unruh thermality in $\mathrm{AdS}$ spacetime \cite{Jennings2010}. In this scheme, we are interested in a uniformly accelerated UDW battery in $\mathrm{dS}_4$ and in $\mathrm{AdS}_4$.

With respect to the uniformly accelerated battery in $\mathrm{dS}_4$, the battery suffers from both the Unruh effect and the Gibbons-Hawking effect. According to the results of \cite{Du2021}, we can get the response per unit time in terms of the temperature $T=\dfrac {\sqrt{a^2+k^2}}{2\pi}$,
\begin{equation}
\label{eq:(14)}
\mathcal{G}^{\mathrm{dS}}(\Omega)=\dfrac {\Omega}{4\pi^2 \ell T(e^{\Omega/T}-1)}.
\end{equation}
In the limit of $a\rightarrow 0$, the temperature $T\rightarrow \frac 1{2\pi \ell}$, which approaches the Gibbons-Hawking temperature for the comoving observer.

For $\mathrm{AdS}_4$, the UDW battery also experiences a nonzero temperature for a uniformly accelerated observer in the condition of the supercritical trajectories \cite{Jennings2010}. We can write the response per unit time in terms of the temperature $T=\dfrac {\sqrt{a^2-k^2}}{2\pi}$,
\begin{equation}
\label{eq:(15)}
\mathcal{G}^{\mathrm{AdS}}(\Omega)=\left(\frac {\Omega}{2\pi}-\dfrac {\zeta}{4\pi \ell\sqrt{4\pi^2T^2\ell^2+1}} \sin \left[\frac {\Omega}{\pi T}\sinh^{-1}(2\pi T \ell) \right]\right)\dfrac {\Theta(T)}{e^{\Omega/T}-1},
\end{equation}
where $\Theta(\cdot)$ is the Heaviside function. The first term represents a thermal spectrum with the temperature $T$, which satisfies the relation of $a>k$. The accelerated detector will witness no Unruh response until the acceleration exceeds the critical curvature scale of $\mathrm{AdS}_4$ spacetime. The second term gives a correction to the thermal spectrum, due to the presence of boundary condtions.

As we will see below, both the cosmological horizon and the uniform acceleration play a fundamental role in the thermodynamics of quantum work extraction for the UDW battery through the response per unit time. In the following, we will study the behaviors of the ergotropy for the UDW battery moving in curved spacetimes in order to reveal the characteristics of vacuum structures in dS and AdS spacetimes.

\section{Dynamics of the UDW battery in curved spacetimes}

In the condition of weak couplings, the initial state of combined system can be approximated as $\rho_{tot}(0)=|g\rangle\otimes |0\rangle \langle 0|$, where $\rho(0)$ is the initial state of the atom and $|0\rangle$ denotes the vacuum state of a massless scalar field in (A)dS spacetime. The evolution $\rho_{tot}$ is driven by the total Hamiltonian given in Eq. (5). From the viewpoint of open quantum systems \cite{Petruccione2002,Benatti2004}, the scalar vacuum field is considered as the environment. Accordingly, the dynamics of the battery alone is obtained by tracing over the environmental degrees of freedom. As rigorously proven in \cite{Davies1974,Davies1976markovian}, under the weak-coupling limit between the battery and the environment, the reduced density matrix $\rho(\tau)=Tr_{f}[\rho_{tot}(\tau)]$ in the interaction picture satisfies,
\begin{equation}
	\label{eq:(16)}
    \frac{\partial \rho(\tau)}{\partial \tau}=-\int_0^\infty  {ds} Tr_f\{[H_I(\tau),[H_I(\tau-s),\rho(\tau)  \otimes |0\rangle \langle 0|]]\}.
\end{equation}
The Born-Markov approximation is considered here \cite{Petruccione2002}. The parameter $f$ denotes the degrees of freedom for the scalar vacuum field in curved spacetimes. The interaction Hamiltonian is $H_I(\tau)=e^{i(H_b+H_{\phi}) \tau}H_Ie^{-i (H_b+H_{\phi}) \tau}=\sigma_1(\tau)\otimes \Phi(\tau)$ where $\sigma_1(\tau)=e^{iH_b \tau}\sigma_1 e^{-iH_b \tau}$ and $\Phi(\tau)=e^{iH_{\phi} \tau}\Phi e^{-iH_{\phi} \tau}$ are the time dependent operators in the interaction picture.

To express the interaction Hamiltonian, we can obtain $e^{iH_b \tau}=e^{i\frac {\Omega}2  \tau}P_{+}+e^{-i\frac {\Omega}2 \tau}P_{-}$ where $P_{\pm}=\frac{1}{2}(I\pm\sigma_1)$ are the two projection operators. Therefore, the operator $\sigma_1(\tau)$ is written in the form of $\sigma_{1}(t)=\sum_{\xi=0, \pm} e^{i \xi \Omega \tau} \sigma_{1}^{(\xi)}$
where $\sigma_{1}^{(0)}=\sum_{k=\pm}P_{k} \sigma_{1} P_{k}= \sigma_1, \;\;\sigma_{1}^{( \pm)}=P_{ \pm} \sigma_{1} P_{\mp}=0$.

The Eq. (16) can be expanded as
\begin{align}
   	\label{eq:(17)}
   	\frac{\partial \rho(\tau)}{\partial \tau} = &\int_0^\infty  ds[\sigma_1(\tau)\rho(\tau)\sigma_1(\tau-s)\otimes\langle 0 |\Phi(\tau) \Phi(\tau-s)|0\rangle \nonumber\\
     & -\sigma_1(\tau)\sigma_1(\tau-s)\rho(t)\otimes\langle 0 |\Phi(\tau) \Phi(\tau-s)|0\rangle + h.c.]
\end{align}
where $\langle 0 |\Phi(\tau) \Phi(\tau-s)|0\rangle=Tr_f[\Phi(\tau) \Phi(\tau-s) |0\rangle \langle 0|]$ is the correlation function of the environment. According to \cite{Petruccione2002}, the reservoir correlation function is homogeneous in time which yields $\langle 0 |\Phi(\tau) \Phi(\tau-s)|0\rangle=\langle 0| \Phi(s) \Phi |0\rangle $. For the curved spacetimes, this term represents the Wightman function for the vacuum state. We consider the rotating wave approximation which neglect the rapidly oscillating terms. Therefore, the part $\sigma_1(\tau)\rho(\tau)\sigma_1(\tau-s)=\sum_{\xi, \xi^{\prime}=0, \pm}e^{i(\xi+\xi^{\prime})\Omega \tau} e^{-i\xi^{\prime}\Omega s} \sigma_1^{(\xi)}\rho(\tau)\sigma_1^{(\xi^{\prime})}$ in Eq. (17) can be approximately expressed as $\sum_{\xi=0, \pm}e^{i\xi \Omega s} \sigma_1^{(\xi)}\rho(\tau)\sigma_1^{(-\xi)}$ where the rotating wave approximation of $\xi^{\prime}+\xi=0$ is applied.

This allows us to explicitly evaluate Eq. (17) and express the result in terms of the Fourier and Hilbert transforms of the environmental correlations, $ \alpha^{(\xi)}(\Omega)=\int_{-\infty}^{\infty} ds e^{i \xi \Omega s}\langle 0|\Phi(s) \Phi| 0\rangle$ and $\beta^{(\xi)}(\Omega)=\int_{0}^{\infty} ds e^{i \xi \Omega s}\langle 0|\Phi(s) \Phi| 0\rangle-\int_{0}^{\infty} ds e^{-i \xi \Omega \tau}\langle 0|\Phi \Phi(s)| 0\rangle$.

Explicitly, we find,
\begin{align}
  	\label{eq:(18)}
      \frac{\partial \rho(t)}{\partial t}= & \frac{1}{2} \sum_{\xi=0, \pm} \left\{\alpha^{(\xi)}\left(2 \sigma_{1}^{(-\xi)} \rho \sigma_{1}^{(\xi)}-\sigma_{1}^{(\xi)} \sigma_{1}^{(-\xi)} \rho-\rho \sigma_{1}^{(\xi)} \sigma_{1}^{(-\xi)}\right)\right.  \left.+\beta^{(\xi)}\left[\rho, \sigma_{1}^{(\xi)} \sigma_{1}^{(-\xi)}\right]\right\}.
\end{align}
By using Eqs. (16)-(18), we can obtain the quantum master equation in the Kossakowski-Lindblad form of,

\begin{align}
\label{eq:(19)}
\frac {\partial}{\partial \tau} \rho(\tau)&\;=\;-i[H^{(eff)}_{b}, \rho(\tau)]+\frac 12\sum_{i,j=1}^{3}a_{ij}\mathcal{D}_{ij}[\rho(\tau)], \\
a_{ij}&\;=\;A\delta_{ij}-iB\varepsilon_{ijk}\delta_{k1}+C\delta_{i1}\delta_{j1}, \nonumber \\
A&\;=\; \frac {\lambda^2}{2}[\mathcal{G}(\Omega)+\mathcal{G}(-\Omega)],\;B= \frac {\lambda^2}{2}[\mathcal{G}(\Omega)-\mathcal{G}(-\Omega)],\;C=\lambda^2 \mathcal{G}(0)-A,\nonumber
\end{align}
where the dissipator $\mathcal{D}_{ij}(\rho)=2\sigma_j\rho\sigma_i-\sigma_i\sigma_j\rho-\rho\sigma_i\sigma_j$ arises from the dissipation and decoherence induced by the environment. The response function $\mathcal{G}(\pm \Omega)=\alpha^{(\pm)}(\Omega)$ is obtained by Eq. (14) or Eq. (15). The Hilbert transform of the Wightman function is expressed as $\mathcal{K}(\pm\Omega)=\beta^{(\pm)}(\Omega)$. In general, $\mathcal{K}(\Omega)=\frac {\mathcal{P}}{\pi i}\int_{-\infty}^{\infty} \mathrm{d}\omega\frac {\mathcal{G}(\omega)}{\omega-\Omega}$ where $\mathcal{P}$ denotes the principle value. The effective Hamiltonian is given by $H^{(eff)}_{b}=\frac {1}{2}\Omega^{'}(\sigma^{+}+\sigma^{-})$ with $\Omega^{'}=\Omega+i\lambda^2[\mathcal{K}(-\Omega)-\mathcal{K}(\Omega)]$ representing the effective coupling. The interaction with external scalar field would have an effect on the Lamb shift \cite{Benatti2004}. In the case of weak couplings, $\lambda^2 \ll \Omega$, we can neglect the Lamb shift in the following.

With respect to a uniformly accelerated battery in (A)dS spacetime, we notice that the field Wightman function fulfills the KMS condition, i.e., $\mathcal{G}^{\mathrm{(A)dS}}(\Omega) = e^{-\frac {\Omega}{T}}\mathcal{G}^{\mathrm{(A)dS}}(-\Omega)$ where $T$ represents the Unruh temperatute. To proceed, we should explore the dynamics of the UDW battery. Therefore, the dynamics of the UDW battery will satisfy the Bloch equation,
\begin{equation}
\label{eq:(20)}
\frac {\mathrm{d}}{\mathrm{d}\tau}\mathbf{r}(\tau)=-2\mathcal{H}\cdot \mathbf{r}(\tau)+\mathbf{\chi},
\end{equation}
where the decaying matrix
\begin{equation}
\label{eq:(21)}
\mathcal{H}=\left(\begin{array}{ccc} 2A & 0 & 0 \\ 0 & 2A+C & \Omega/2 \\ 0 & -\Omega/2 & 2A+C  \end{array}  \right).
\end{equation}
and $\mathbf{\chi}=(-4B, 0, 0)^{\mathrm{T}}$ is the inhomogenous vector. We use a quantum channel to describe the dynamics of the battery by mapping the Bloch vector,
\begin{equation}
\label{eq:(22)}
\mathbf{r}(\tau)=\mathbf{\Gamma}(\tau)\cdot \mathbf{r}(0)+\mathbf{\Lambda}(\tau),
\end{equation}
where \begin{equation}
\mathbf{\Gamma}(\tau) = \exp(-2\mathcal{H}\tau)=\left(\begin{array}{ccc} e^{-4A\tau} & 0 & 0 \\ 0 & e^{-2(2A+C)\tau}\cos \Omega \tau & -e^{-2(2A+C)\tau}\sin \Omega \tau \\ 0 & e^{-2(2A+C)\tau}\sin \Omega \tau & e^{-2(2A+C)\tau}\cos \Omega \tau  \end{array}  \right) \nonumber
\end{equation}
denotes the mapping matrix of the quantum channel and $\mathbf{\Lambda}(\tau) = \frac 12[\mathbf{I}-\mathbf{\Gamma}(\tau)]\mathcal{H}^{-1}\cdot \mathbf{\chi}$ is the mapping vector.

When the ground state is chosen as the initial state $|\psi(0)\rangle  = |g\rangle$, the expression of the Bloch vector for the UDW battery is obtained,
\begin{equation}
\label{eq:(23)}
\mathbf{r}(\tau ) = \left( {\begin{array}{*{20}{c}}
		{{\gamma }{e^{ - 4A\tau }} - \gamma}\\
		{ {e^{ - 2(2A + C)\tau }}\sin \Omega \tau  }\\
		{-{e^{ - 2(2A + C)\tau }}\cos \Omega \tau  }
\end{array}} \right),
\end{equation}
where the ratio $\gamma$ is
\begin{equation}
\gamma=\gamma^{\mathrm{(A)dS}}=-\tanh\left( \frac {\Omega}{2T} \right),  \nonumber
\end{equation}
determined by the temperature $T$ due to the KMS condition. It is found that in the asymptotic limit of $\tau \rightarrow \infty$, the state of the battery arrives at the steady state of $\mathbf{r}_s=-\gamma(1,0,0)^{\mathrm{T}}$ which is related to the thermalization in curved spacetimes.

This method of the open quantum system is more general than simply estimating the transition probability, since it allows us to characterize the dynamics of the state of the UDW battery. By using this approach, we also avoid troubles of regularization. Next, we will investigate quantum work extraction for probing the thermal nature and vacuum structures in dS and AdS.

\section{Thermodynamical behaviors of ergotropy}

This study aims to explore quantum work extraction of the UDW battery described in Eq. (4), which is used to witness the thermality in curved spacetime. Since the steady state depends only on the temperature, we expect the asymptotic behavior of quantum work extraction to inherently reveal the global thermal nature, in accordance with the KMS condition. We propose that the various evolutions of the ergotropy can highlight the discriminations of vacuum structures in dS and AdS spacetimes, which stem from the responses of the UDW battery interacting with a scalar background. Moreover, we will examine how to effectively amplify the ergotropy by properly varying the boundary conditions and spacetime dimensionality.

We emphasize that the UDW battery in our scheme is not only charged by the external driving field, but also by the coupling to the scalar vacuum field in curved spacetimes. With no external driving field, $\Omega=0$, the energy of the battery is just supplied by the thermality and then the asymptotic ergotropy is null because of $r(\infty)=-r_3(\infty)$ determined by the thermal equilibrium state. In the following study, we will work with dimensionless parameters by rescaling the acceleration and the proper time as
\begin{equation}
\label{eq:(24)}
\tilde{a}=\frac {a}{\Omega}, \;\tilde{\tau}=\frac {\lambda^2 \Omega \tau}{2\pi}.
\end{equation}
For convenience, we continue to term $\tilde a $ and $\tilde \tau$ as $a$ and $\tau$, respectively.

After evolving for enough long time, the ergotropy in the asymptotic limit of $\tau\rightarrow \infty$ is obtained as
\begin{equation}
\label{eq:(25)}
\xi^{\mathrm{(A)dS}}_s=\frac{|\gamma^{\mathrm{(A)dS}}|}2.
\end{equation}
As depicted in Fig. 1, it is shown that the asymptotic ergotropy is dependent on the acceleration. It shows that the ergotropy gradually diminishes when the acceleration is increased. The higher charging performance of the UDW battery can be achieved in the condition of the smaller acceleration. It is reasonable that quantum decoherence at a low accelerations has weak impacts on quantum work extraction. This result indicates that quantum work extraction primarily captures the global feature of acceleration-induced Unruh thermalization in two curved $\mathrm{(A)dS}_4$ spacetimes. We find that the condition of $a>k$ guarantees the presence of the asymptotic ergotropy in $\mathrm{AdS}_4$ spacetime. When the acceleration of the battery is larger than the curvature scale of $\mathrm{AdS}_4$, the steady-state ergotropy can be used to estimate the thermal spectrum obeying the KMS condition. Furthermore, the ergotropy of $\mathrm{AdS}_4$ is higher than that of $\mathrm{dS}_4$ at the same acceleration. The asymptotic ergotropy in $\mathrm{AdS}_4$ spacetime decays more slowly than in $\mathrm{dS}_4$ spacetime.

Besides, we pay attention to the transient dynamics of the ergotropy for the UDW battery in two different curved spacetimes. According to the expression of Eq. (4), we can write the time-dependent ergotropy in the form of
\begin{equation}
\label{eq:(26)}
\xi^{\mathrm{(A)dS}}(\tau)=\frac 12 e^{-2(2A+C)\tau}\left[ \sqrt{1+\gamma^2 \delta^2}-\cos \Omega \tau \right].
\end{equation}
Here, the ratio $\gamma=\gamma^{\mathrm{(A)dS}}$ and the symbol $\delta=e^{2C\tau}(e^{4A\tau}-1)$ are determined by the response $\mathcal{G}^{\mathrm{(A)dS}}(\Omega)$ of the UDW battery in curved spacetimes. To characterize the vacuum structures of (A)dS spacetimes, we study the local behaviors of the ergotropy when the curvature and boundary conditions are changed.

Figure 2(a) depicts the evolution of the ergotropy in $\mathrm{dS}_4$ with a positive curvature. It is seen that the oscillatory behavior of the ergotropy occurs with time, and gradually reaches a certain steady value. At the initial stage, quantum work extraction is rapidly enhanced to one maximal value. For a short charging time, the inhibitory effect of acceleration on ergotropy is so weak. When the charging time is increased, the decay of the ergotropy with the acceleration is more apparent. The various behaviors of the ergotropy for $\mathrm{AdS}_4$ spacetime with a negative curvature are shown in Figs. 2(b)-2(d). Three kinds of boundary conditions $\zeta=1,0,-1$ are considered respectively. In contrast to $\mathrm{dS}_4$, a significant decrease in the ergotropy with the acceleration is observed in $\mathrm{dS}_4$ for a short charging time. At a large acceleration, the ergotropy can be quickly suppressed to a steady value. The oscillation effect of the ergotropy is strongest for $\zeta=1$ Dirichlet boundary condition, weaker for $\zeta=0$ transparent boundary condition, and weakest for $\zeta=-1$ Neumann boundary condition. The corrections to quantum work extraction due to various boundary conditions become smaller when the accelerations are larger.

Our previous work has proven the effect of the spacetime dimensionality on Unruh thermality in a $D-$ dimensional Minkowski flat spacetime \cite{Chen2025}. It is natural to explore how the dimensionality of curved spacetime affects ergotropy. Here we take into account a $D-$ dimensional anti-de Sitter spacetime. According to the result of \cite{Jennings2010}, we can obtain the Wightman function along the supercritical acceleration trajectory in terms of
\begin{align}
\label{eq:(27)}
G^{\mathrm{AdS}}_D(\Delta \tau)= & \dfrac {(2\pi T)^{D-2} \Gamma(\frac D2 -1)} {i^{D-2}(4\pi)^{\frac D2}} \left[ \dfrac {1}{\sinh^{D-2} \left(\frac {2\pi T \Delta \tau} {2}-i\epsilon  \right)} \right] \nonumber \\
                                 &-\dfrac {(2\pi T)^{D-2} \Gamma(\frac D2 -1)} {(4\pi)^{\frac D2}} \left[ \prod_{j=0,1} \dfrac {1}{\left( \sinh(\tilde{k}+(-1)^j(\frac {2\pi T \Delta \tau}{2}-i\epsilon)) \right)^{\frac D2 -1} }\right],
\end{align}
where the parameter $\tilde{k}=\sinh^{-1}(\frac {2\pi T}{k})$ and $T=\dfrac {\sqrt{a^2-k^2}}{2\pi}=\frac 1{\beta}$ denotes the observed temperature in $\mathrm{AdS}_D$ spacetime. $\Gamma(\cdot)$ represents the Gamma function. The second term of Eq. (27) is purely from the nonzero curvature of $\mathrm{AdS}_D$ spacetime. For odd dimensions, the evaluation of the second term becomes more awkward to work with \cite{Jennings2010}. However, we can still analyze thermality from the fact that $G^{\mathrm{AdS}}_D(\Delta \tau+i \beta)=(-1)^D G^{\mathrm{AdS}}_D(\Delta \tau)$, which is the KMS condition of thermality for a fermionic field \cite{Ooguri1986}. For example, we can derive the analytical expression of the response function for $D=6$,
\begin{equation}
\label{eq:(28)}
\mathcal{G}^{\mathrm{AdS}}_{6}(\Omega)=\left(\frac {\Omega^3+\Omega T^2}{3}-\frac {k^4}{16\pi^2 a^2} \left [ \frac {2a^2-k^2}{2a}\sin (\frac {\Omega \tilde{k}}{\pi T}) -\Omega \cos (\frac {\Omega \tilde{k}}{\pi T})  \right]   \right)\dfrac {\Theta(a-k)}{e^{\Omega/T}-1},
\end{equation}
where the Dirichlet boundary condition $\zeta=1$ is considered. By using the response for $D=4,6$ in AdS spacetime, we numerically investigate the evolution of the ergotropy. From Figure 3, we find that the increase of the dimension tends to enhance the ergotropy of the UDW battery at the initial stage of the evolution. This is the reason that the response rate is increased with spacetime dimension \cite{Jennings2010}. The fast thermalization can be demonstrated by the rapid stabilization of oscillatory behavior of the ergotropy in a high-dimensional AdS spacetime. However, for any dimension, the asymptotic ergotropy keeps the same one like Eq. (25), which is mainly determined by the acceleration and curvature.

\section{Discussion}
We have investigated the ability of the accelerated Unruh-DeWitt battery to probe thermality and vacuum structures in both de Sitter and anti-de Sitter spacetimes. By analyzing the response functions, we have identified some key factors affecting the thermodynamics of quantum work extraction of the relativistic battery in these curved spacetime backgrounds. Several interesting results are obtained as:

(i) In both dS and AdS spacetimes, the UDW battery eventually thermalizes with the environment due to long-term couplings, leading to the steady values of quantum work extraction in the asymptotic limit. The asymptotic ergotropy is regarded as an effective witness to the thermality with the acceleration and spacetime curvature. It is interesting to find that the asymptotic ergotropy will approach the same one when the condition of $a\gg k$ is satisfied. To be explicit, the lower temperature, the greater asymptotic ergotropy. The steady-state value of the ergotropy is independent of the spacetime dimension, which is determined by the KMS condition. In addition, the asymptotic ergotropy in AdS is larger than that in dS at a certain acceleration.

(ii) The effect of the acceleration on the dynamics of the ergotropy in both dS and AdS is prominent. The increase of the acceleration will lead to the thermal noise with high temperatures. The evolutions of quantum work extraction in curved spacetimes exhibit oscillatory behaviors at low accelerations. Furthermore, this behavior can be suppressed for higher accelerations. It is revealed that the oscillating phenomenon of the ergotropy in dS spacetime is significant during the evolution. However, it is observed that as the acceleration increases, the influence of the acceleration gradually weakens until it disappears. When various kinds of boundary conditions in AdS are considered, the Dirichlet condition contributes to the noticeable enhancement of the ergotropy at low accelerations.

(iii) There are quite different sensitivities of the UDW battery for the change of the acceleration in two curved spacetimes. In AdS, the UDW battery is more sensitive to the acceleration-dependent temperature than that in dS. This means that the UDW battery in AdS has already thermalized at the same acceleration, thereby eliminating the influence of other factors such as different boundary conditions on quantum work extraction. We also find that the influence of the acceleration on the ergotropy is very weak in the early stage of the evolution. This is the reason that the thermal effect induced by vacuum fluctuations is so weak that the ergotropy for a very short charging time is mainly determined by the external driving coherent field.

(iv) From the viewpoint of energy storage, the ergotropy related to the response function can also demonstrate the emergence of statistic inversion in odd dimensional curved AdS spacetime. This anomalous inversion of statistics could not destroy the thermal nature of Unruh effect, which is only dependent on the KMS condition. It turns out that the increase of the dimension of AdS spacetime can improve the ergotropy at a certain extent. The fast thermalization of the UDW battery has be proven in the higher dimension AdS spacetime. In the asymptotic limit, the ergotropy for any dimensional AdS will reach a constant value which is determined by the acceleration and spacetime curvature.

With all these in mind, we demonstrate that the current observations from the relativistic UDW battery can offer deeper insights into the study of vacuum structures in de Sitter and anti-de Sitter spacetimes. The present results can help achieve relativistic quantum information tasks in the frame of curved spacetimes.

\begin{acknowledgments}
We would like to thank Professor Bill Unruh and Professor Philip C. E. Stamp for the discussions on the related work. This work is supported by $\mathrm{SCOAP}^{3}$.
\end{acknowledgments}

\newpage

{\large \bf Figure Captions}

\vskip 0.5cm

{\bf Figure 1.}

The asymptotic behaviour of the ergotropy of an accelerated UDW battery in $\mathrm{(A)dS}_4$ spacetimes, which is a function of the acceleration. The global Unruh thermal nature in curved spacetime is shown by the monotonic decrease behavior. The curvature scale $k=1$ in $\mathrm{(A)dS}_4$ spacetimes is chosen. The red solid line represents the case of AdS and the black dashed line denotes that of dS.

\vskip 0.5cm

{\bf Figure 2.}

The evolution of the ergotropy for the UDW battery in $\mathrm{(A)dS}_4$ sapcetime, which is a function of scaled proper time $\tau$ and acceleration $a$ in the condition of curvature $k=1$. (a) The ergotropy for $\mathrm{dS}_4$. (b) The ergotropy for $\mathrm{AdS}_4$ with the Dirichlet boundary condition $\zeta=1$. (c) The ergotropy for $\mathrm{AdS}_4$ with the transparent boundary condition $\zeta=0$. (d) The ergotropy for $\mathrm{AdS}_4$ with the Neumann boundary condition $\zeta=-1$.

\vskip 0.5cm

{\bf Figure 3.}

The oscillatory behaviors of the ergotropy for the UDW battery in $D=4,6$ dimensional $\mathrm{AdS}_D$ spacetime. The black solid line represents the case of $D=4$ and the red dashed line denotes that of $D=6$ spacetime. The scaled dimensionless acceleration $a=3$ and curvature $k=1$ are chosen.

\newpage
\begin{figure}
  \centering
  \includegraphics[width=1.0\textwidth]{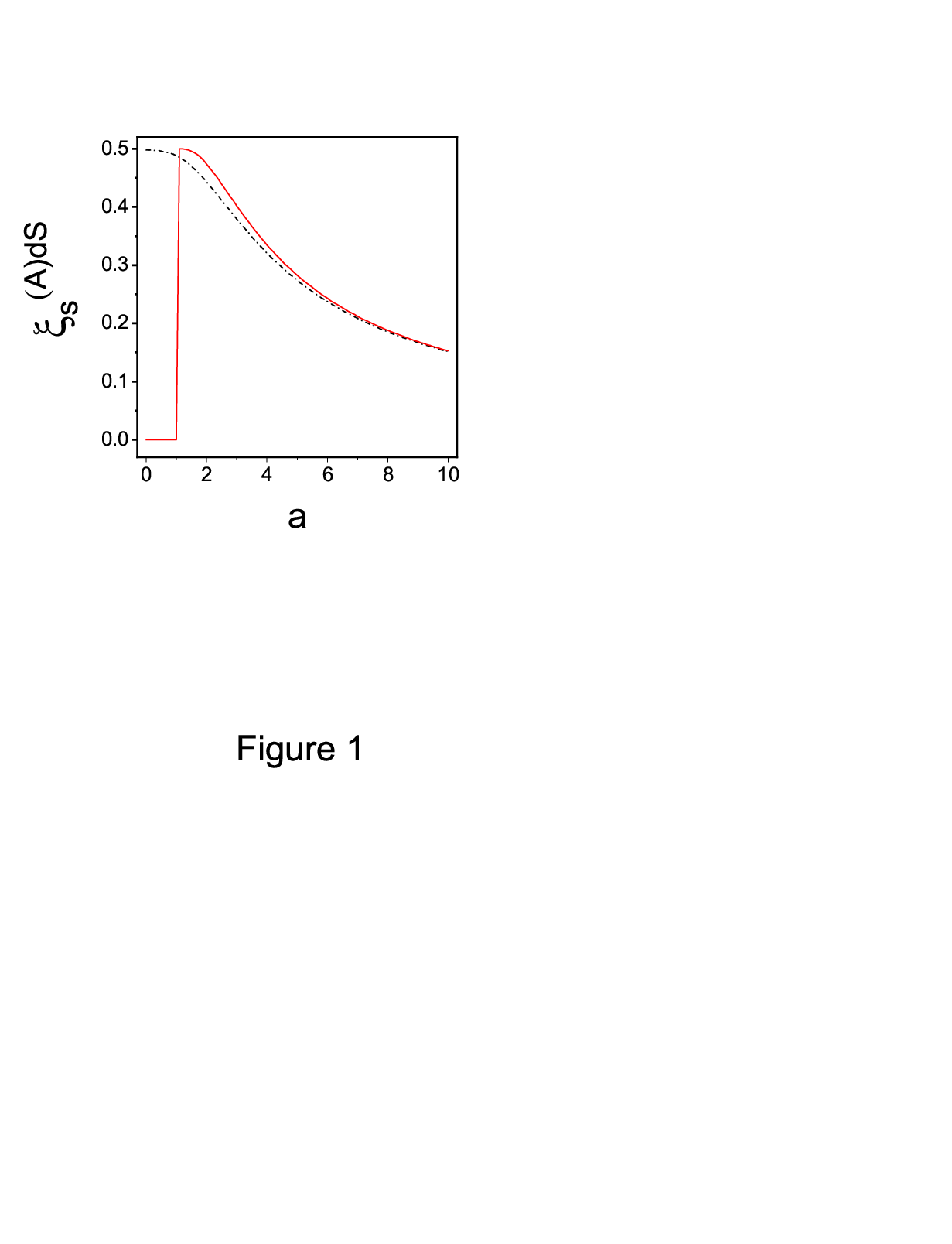}
\end{figure}

\begin{figure}
  \centering
  \includegraphics[width=1.0\textwidth]{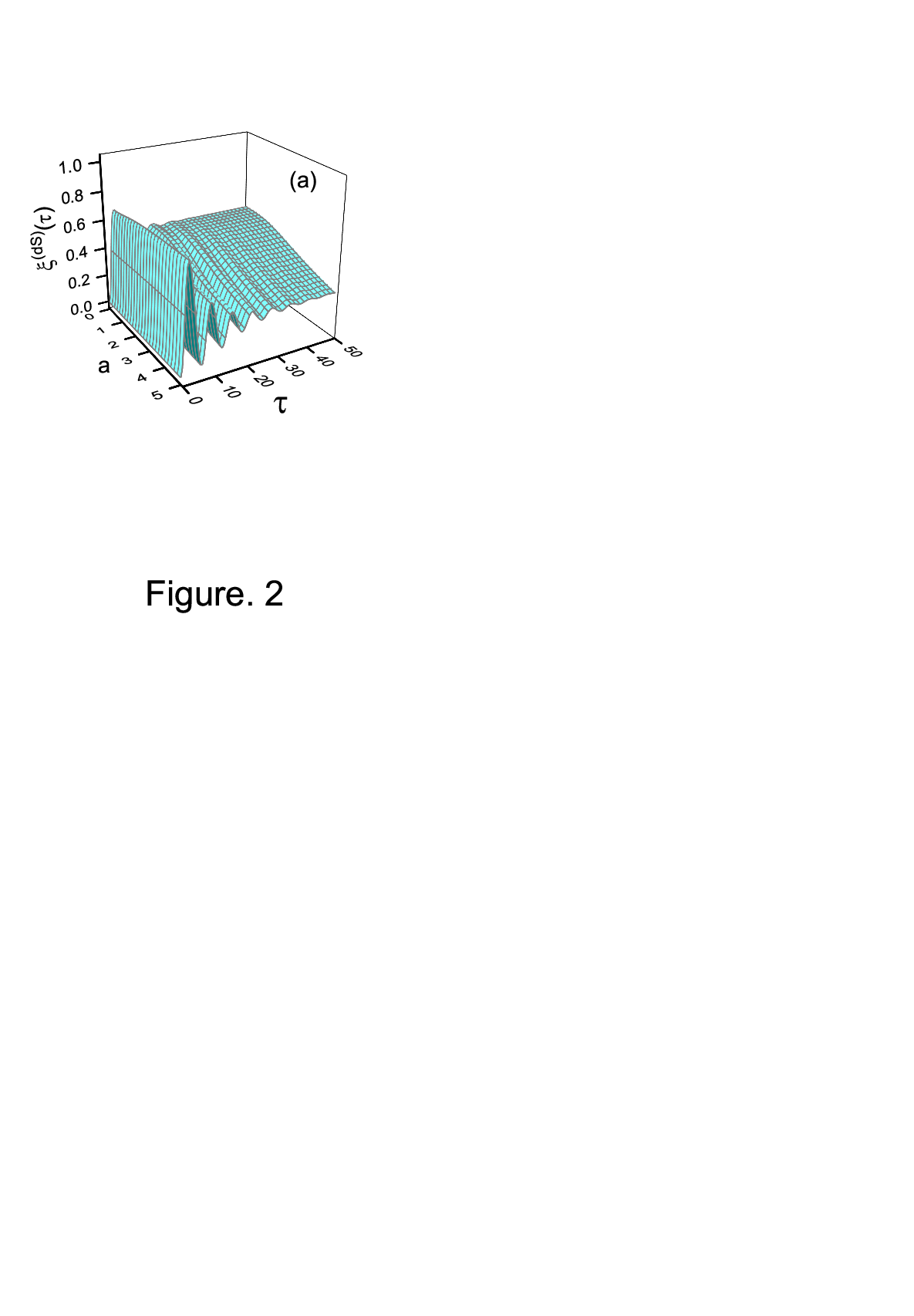}
\end{figure}

\begin{figure}
  \centering
  \includegraphics[width=1.0\textwidth]{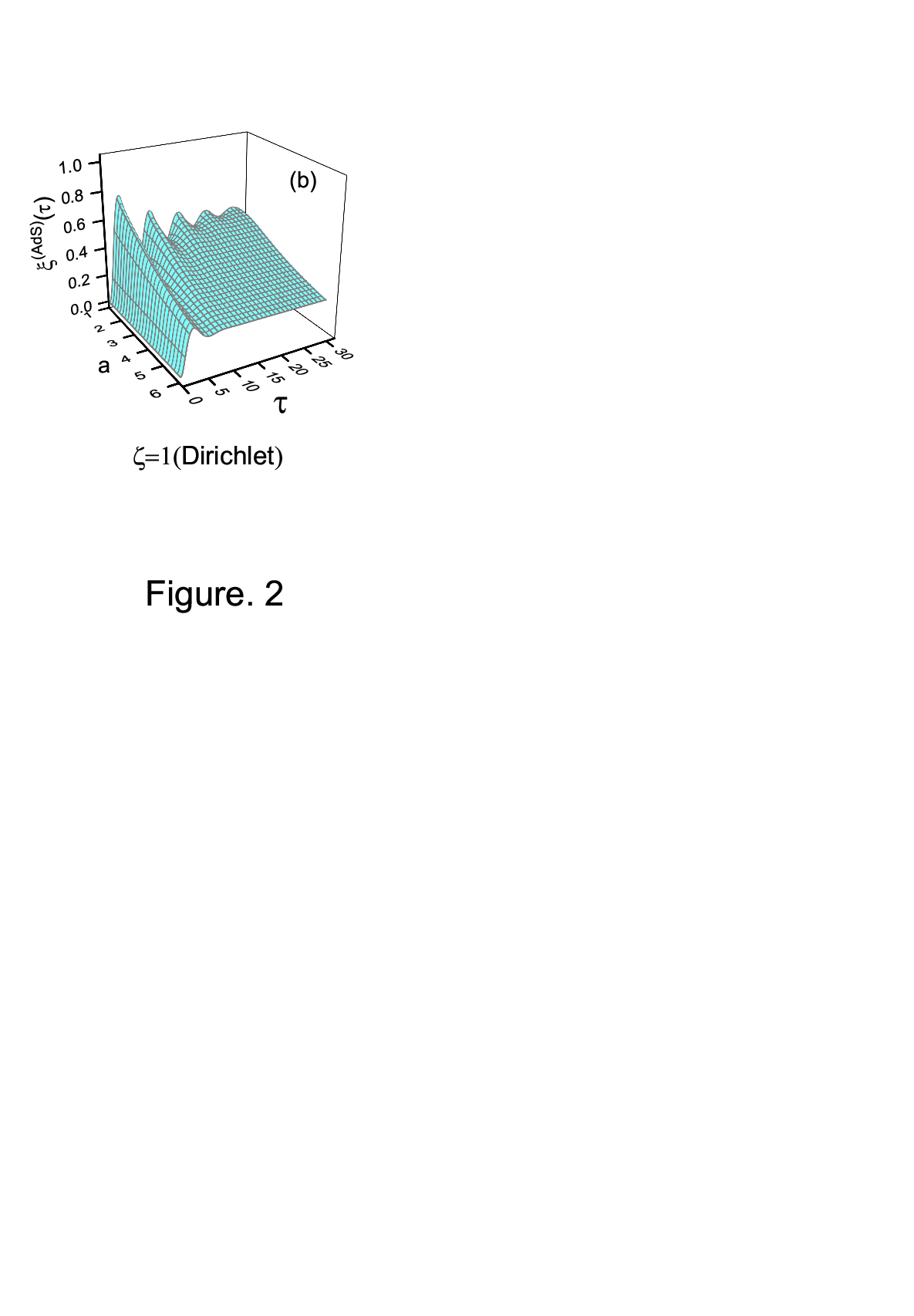}
\end{figure}

\begin{figure}
  \centering
  \includegraphics[width=1.0\textwidth]{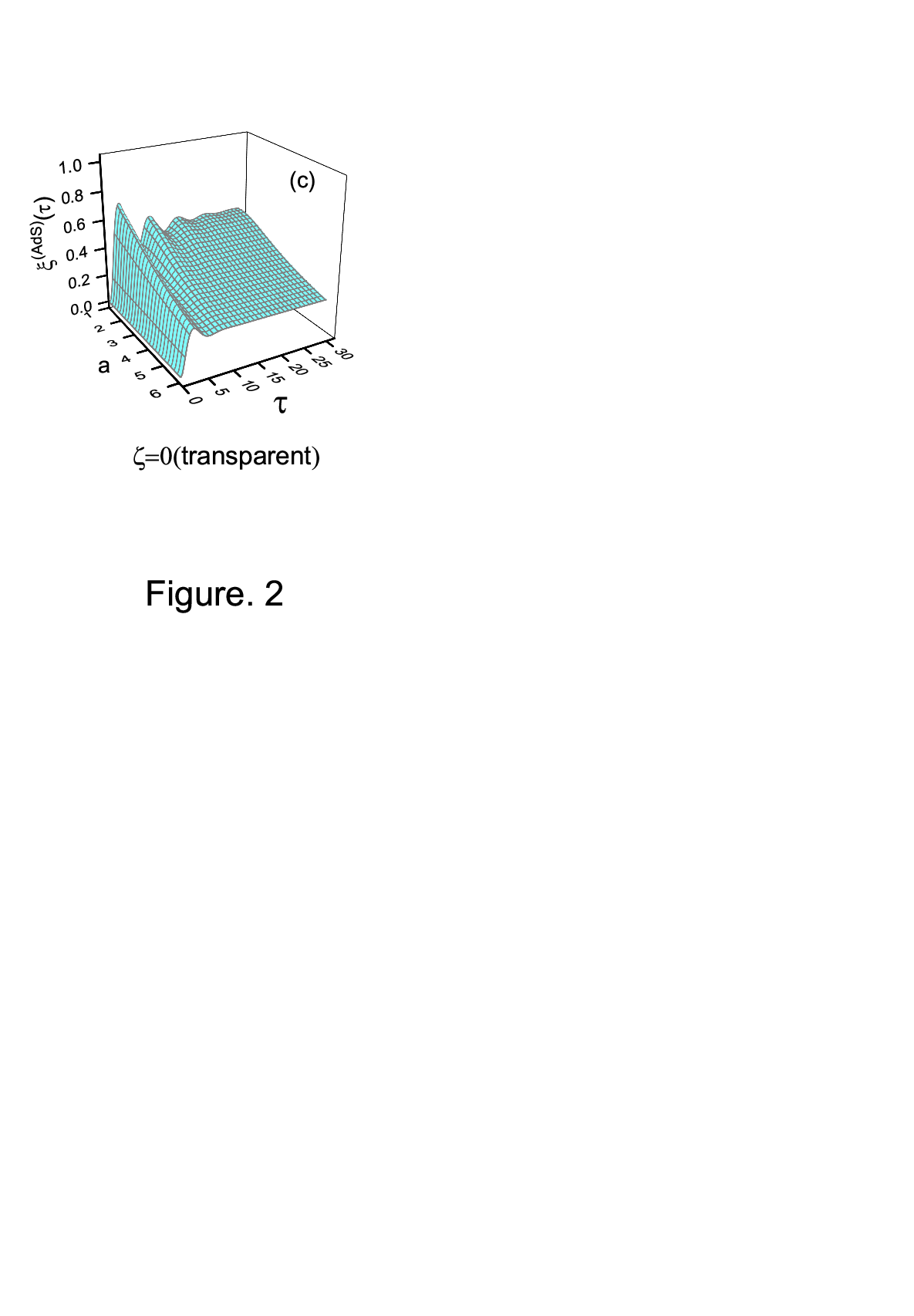}
\end{figure}

\begin{figure}
  \centering
  \includegraphics[width=1.0\textwidth]{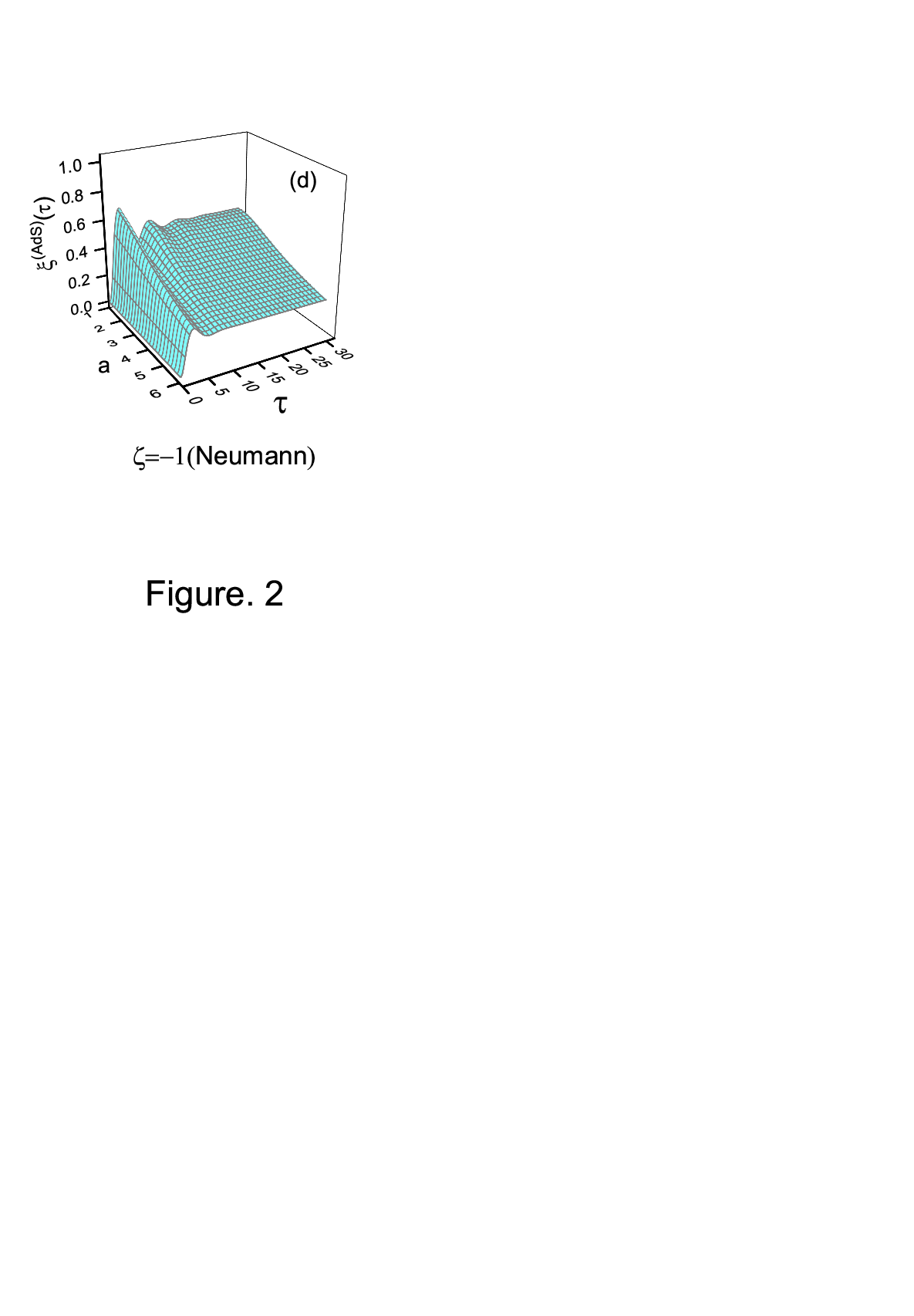}
\end{figure}

\begin{figure}
  \centering
  \includegraphics[width=1.0\textwidth]{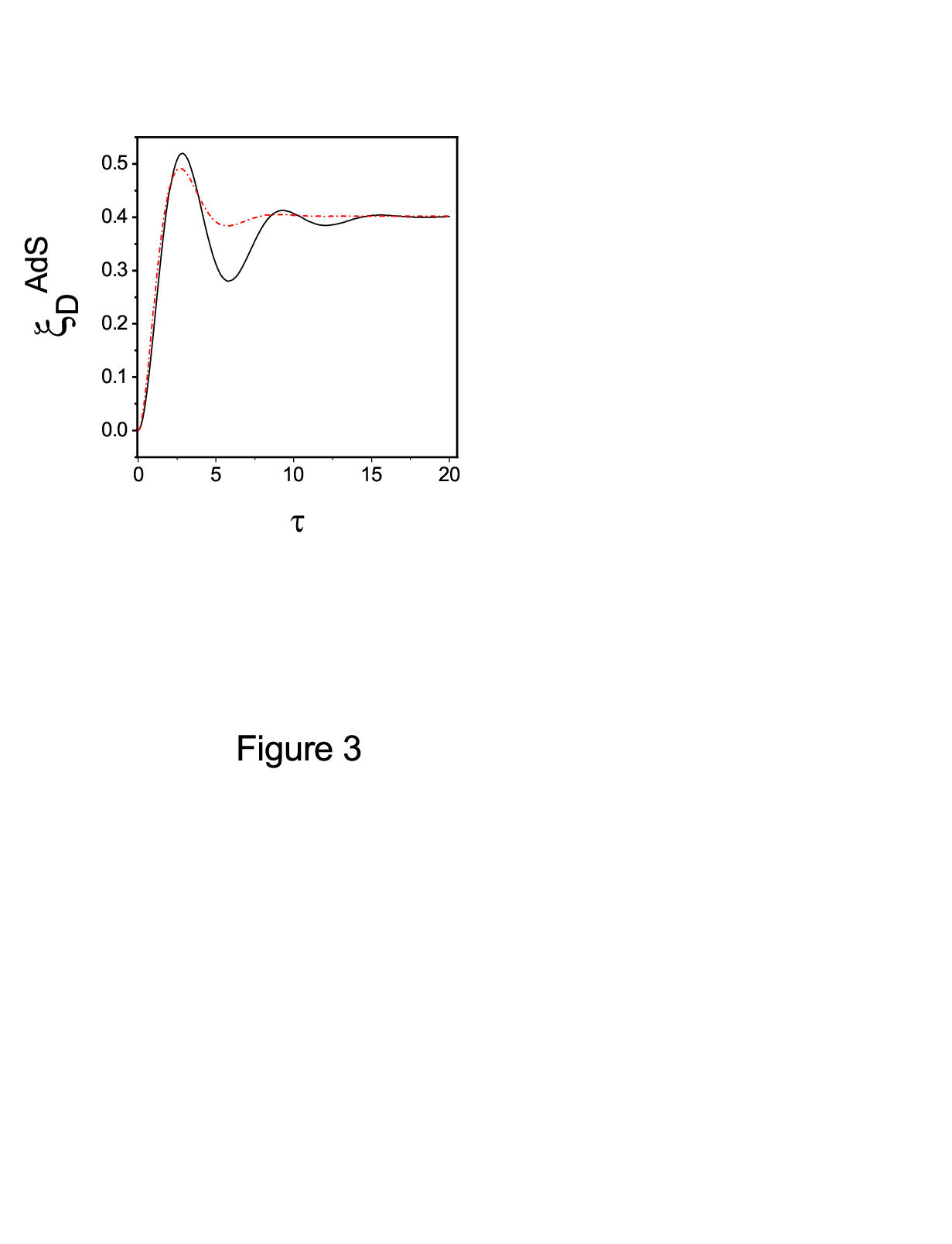}
\end{figure}

\end{document}